\documentclass[10pt]{article}
\usepackage{graphicx}
\begin{document}

\twocolumn[\begin{center}
{\large {\bf Hawking Non-thermal and Thermal Radiations of Schwarzschild Anti-de Sitter Black\\ Hole by Hamilton-Jacobi method}}\\
\vspace{.5cm}
M. Atiqur Rahman\\
{\it Department of Applied Mathematics, Rajshahi University, Rajshahi - 6205, Bangladesh}\\
M. Ilias Hossain\\
{\it Department of Mathematics, Rajshahi University, Rajshahi - 6205, Bangladesh}\\
\end{center}
\centerline{\bf Abstract}
\baselineskip=18pt
\bigskip
\begin{center}
\parbox{16cm}{The massive particles tunneling  method has been used to investigate the Hawking non-thermal and purely thermal radiations of Schwarzschild Anti-de Sitter (SAdS) black hole. Considering the spacetime background to be dynamical, incorporate the self-gravitation effect of the emitted particles the imaginary part of the action has been derived from Hamilton-Jacobi equation. Using the conservation laws of energy and angular momentum we have showed that the non-thermal and purely thermal tunneling rates are related to the change of Bekenstein-Hawking entropy and the derived emission spectrum deviates from the pure thermal spectrum. The result obtained for SAdS black hole is also in accordance with Parikh and Wilczek\rq s opinion and gives a correction to the Hawking radiation of SAdS black hole.
{\bf PACS number(s)}: 04.70.-s, 04.70.Dy, 97.60.Lf\\{\bf E-mail}:atirubd@yahoo.com*, ilias\_math@yahoo.com}
\end{center}
\vspace{0.2cm}
]\section{Introduction}\label{sec1}
According to the information loss paradox \cite{one,two}, the information carried out by a physical system falling toward black hole singularity has no way to recover after a black hole has completely disappeared. The loss of information was considered as preserved inside the black hole and so was not a serious problem in the classical theory. Taking quantum process into account, the situation has changed. With the emission of thermal radiation \cite{one,two}, black holes could lose energy, shrink, and eventually evaporate away completely. In this basis, many research works on the thermal radiation of black holes have been made \cite{three,four,five,six}. Since the radiation with a precise thermal spectrum carries no information, the original matter that forms the black hole can evolve as the thermal spectrum at infinity and violates the fundamental principles of quantum theory because of prescribing a unitary time evolution of basis states. However, the information paradox can perhaps be attributed to the semi-classical nature of the investigations of Hawking radiation. Researches in string theory indeed support the idea that Hawking radiation can be described within a manifestly unitary theory and is still remains a mystery how information is recovered. Although a complete resolution of the information loss paradox might be within a unitary theory of quantum gravity or string/M theory, it is argued that the information could come out if the outgoing radiation were not exactly thermal but had subtle corrections \cite{three}.

A semi-classical method to describe Hawking radiation as tunneling process was initiated by Kraus and Wilczek \cite{seven,eight}, where a particle moves in dynamical geometry. This method involve calculating the imaginary part of the action for the process of s-wave emission across the horizon, which in turn is related to the Boltzmann factor for emission at the Hawking temperature. Applying this method, two different methods have been employed to calculate the imaginary part of the action, one the null geodesic method developed by Parikh and Wilczek \cite{nine,ten,eleven} and other by Angheben et al. \cite{twelve}. In fact, the method of Angheben et al. \cite{twelve} is an extension of the complex path analysis proposed by Padmanabhan et al. \cite{thirteen,fourteen,fifteen}. The latter method involves consideration of a emitted scalar particle, ignoring its self-gravitation and assumes that its action satisfies the relativistic Hamilton-Jacobi equation. An appropriate ansatz for the action can be obtained from the symmetries of the spacetime which is known as the Hamilton-Jacobi ansatz. Both the methods show that when the self-gravitational interaction and the unfixed background spacetime are taken into account, the actual Hawking radiation spectrum deviates from the purely thermal one, satisfies the underlying unitary theory and gives a leading correction to the radiation spectrum. Extending these methods to general case, a lot of works for various spacetimes have been done \cite{sixteen,seventeen,eighteen,nineteen,twenty,twenty one,twenty two,twenty three,twenty four,twenty five,twenty six,twenty seven,twenty eight,twenty nine,thirty,thirty one,thirty two,thirty three,thirty four,thirty five,thirty six,thirty seven,thirty eight,thirty nine,fourty,fourty one} and all of these are limited to massless particle.

On the other hand, Hawking radiation from massive uncharged particle tunneling \cite{fourty two} and charged particle tunneling \cite{fourty three,fourty four} from black hole was first proposed by Zhang and Zhao. Exploiting this work, a few researches have been carried out as charged particle tunneling \cite{fourty five,fourty six,fourty seven}. Recently, Kerner and Mann developed quantum tunneling methods for analyzing the temperature of Taub-NUT black holes \cite{fourty eight} using both the null-geodesic and Hamilton-Jacobi methods. In the latter method the self-gravitation interaction and energy conservation of emitted particle were ignored to calculate the thermal radiation spectrum. Parikh and Wilczek have shown that these radiation spectrum is not strictly thermal but satisfies the underlying unitary theory when self-gravitation interaction and energy conservation are considered. Considering Kerner and Mann\rq s process Chen, Zu and Yang reformed Hamilton-Jacobi  method for massive particle tunneling and investigate the Hawking radiation of the Taub-NUT black hole \cite{fourty nine}. Using this method Hawking radiation of Kerr-NUT  black hole \cite{fifty}, the charged black hole with a global monopole \cite{fifty one} and Schwarzschild-de Sitter black hole \cite{fifty two} have been reviewed. We apply these method to investigate the Hawking radiation of SAdS black hole.

The solutions of black holes in Anti-de Sitter spaces come from the Einstein equations with a negative cosmological constant. Anti-de Sitter black holes are different from de Sitter black holes. The difference consisting in them is due to minimum temperatures that occur when their sizes are of the order of the characteristic radius of the anti-de Sitter space.  For larger Anti-de Sitter black holes, their red-shifted temperatures measured at infinity are greater. This implies that such black holes can be in stable equilibrium with thermal  radiation at a certain temperature. Moreover, recent development in string /M-theory greatly stimulate the study of black holes in anti-de Sitter spaces. One example is the AdS/CFT correspondence \cite{fifty three,fifty four,fifty five} between a weakly coupled gravity system in an anti-de Sitter background and a strongly coupled conformal field theory on its boundary. So our study on the Schwarzschild anti-de Sitter black holes is reasonable and meaningful.

This paper is structured as follows. The next section will outline the position of event horizon of SAdS black hole. In section 3, we then consider the unfixed background spacetime and the self-gravitational interaction into account, we review the Hawking non-thermal radiation of SAdS black hole from massive particle tunneling method. The new line element of SAdS black hole near the even horizon is also derived in this section. In section 4, we have derived the Hawking purely thermal radiation from non-thermal rate. Finally, in section 5, we present our remarks.

\section{Schwarzschild Anti-de Sitter\\ black hole}\label{sec2}
The Schwarzschild Anti-de Sitter black hole with mass $M$ and a negative cosmological constant $\Lambda =-3/\ell^2 $ is given by
\begin{equation}
ds^2=-f(r)dt^2+\frac{1}{f(r)}dr^2+r^2(d\theta ^2+{\rm sin}^2\theta d\varphi ^2),\label{eq1}
\end{equation}
where the lapse function, $f(r)$, is given by
\begin{equation}
f(r)=1-\frac{2m}{r}+\frac{r^2}{\ell^2},\label{eq2}
\end{equation}
and the coordinates are defined such that $-\infty\leq t\leq \infty $, $r\geq 0$, $0\leq \theta \leq \pi $ and $0\leq \phi \leq 2\pi$.
The lapse function vanished at the zeros of the cubic equation
\begin{equation}
r^3+\ell^2r-2m\ell^2=0.\label{eq3}
\end{equation}
The only real roots of this equation is
\begin{eqnarray}
r_+=\frac{2}{3}\sqrt{3}\ell\,\sinh\left(\frac{1}{3}\sinh^{-1}\left(3\sqrt{3}\frac{m}{\ell}\right)\right).\label{eq4}
\end{eqnarray}
Expanding $r_+$ in terms of $m$ with $1/\ell^2<<m^2/9$, we obtain
\begin{equation}
r_+=2m\left(1-\frac{4m^2}{\ell^2}+ ....\right).\label{eq5}
\end{equation}
Therefore, we can write $r_+=2m\eta, $ with $\eta <1$. The event horizon of the SAdS black hole is smaller than the Schwarzschild event horizon, $r_H=2m$.

\section{The Hamilton-Jacobi Method\\ for Non-thermal Radiation}\label{sec4}
We next consider the method of Chen et al. \cite{fourty nine} for calculating the imaginary part of the action making use of the Hamilton-Jacobi equation \cite{twelve}. We assume that the action of the outgoing particle is given by the classical action $I$ satisfies the relativistic Hamilton-Jacobi equation
\begin{equation}
g^{\mu\nu}\left(\frac{\partial I}{\partial x^\mu}\right)\left(\frac{\partial I}{\partial x^\nu}\right)+u^2=0,\label{eq6}
\end{equation}
in which $u$ and $g^{\mu\nu}$ are the mass of the particle and the inverse metric tensors derived from the line element (\ref{eq1}). Since the event horizon of SAdS black hole coincides with the outer infinite redshift surface, here we can apply the geometrical optics limit. Using the WKB approximation \cite{fifty six}, the tunneling probability for the classically forbidden trajectory of the s-wave comming from inside to outside of SAdS event horizon is given by
\begin{eqnarray}
\Gamma \sim {\rm exp}(-2{\rm Im}I) \label{eq7}.
\end{eqnarray}

As mention before, this method is different from Parikh and Wilczek method (Null geodesic) in which the action mainly relies on the exploration of the equation of motion in the Painlev\'e coordinates systems and the calculation of Hamilton equation. But in the Hamilton-Jacobi method we avoid this for calculating the imaginary part of the action $I$.
For the convenient of discussion, we define $\Delta=r^2-2mr+\frac{r^4}{\ell^2}$ and then the line element (\ref{eq1}) can be written as
\begin{equation}
ds^2=-\frac{\Delta}{r^2}dt^2+\frac{r^2}{\Delta}dr^2+r^2(d\theta^2+{\rm sin^\theta}d\phi^2)\label{eq8}.
\end{equation}
Near the event horizon, the above line element can be rewritten as
\begin{eqnarray}
ds^2=-\frac{\Delta_{,r}(r_+)(r-r_+)}{r^2_+}dt^2+\frac{r^2_+}{\Delta_{,r}(r_+)(r-r_+)}dr^2\nonumber\\
+r^2_+(d\theta^2+{\rm sin^2\theta}d\phi^2)\label{eq9},
\end{eqnarray}
where,
\begin{equation}
\Delta_{,r}(r_+)=\frac{d\Delta}{dr}\bigg |_{r=r_+}=2(r_+-m+2\frac{r^3_+}{\ell^2}).\label{eq10}
\end{equation}
\onecolumn
For the metric (\ref{eq9}), the non-null inverse metric tensors are
\begin{eqnarray}
&&g^{00}=-\frac{r^2_+}{\Delta_{,r}(r_+)(r-r_+)}, \quad g^{11}=\frac{\Delta_{,r}(r_+)(r-r_+)}{r^2_+}, \nonumber\\
&&g^{22}=\frac{1}{r_+^2}, \quad g^{33}=\frac{1}{r_+^2{\rm sin^2\theta}}.\label{eq11}
\end{eqnarray}
The Hamilton-Jacobi equation (\ref{eq6}), with the help of Eq. (\ref{eq11}) becomes
\begin{equation}
-\frac{r^2_+}{\Delta_{,r}(r_+)(r-r_+)}\left(\frac{\partial I}{\partial t}\right)^2+\frac{\Delta_{,r}(r_+)(r-r_+)}{r^2_+}\left(\frac{\partial I}{\partial r}\right)^2+\frac{1}{r_+^2}\left(\frac{\partial I}{\partial \theta}\right)^2+\frac{1}{r_+^2{\rm sin^2\theta}}\left(\frac{\partial I}{\partial \phi}\right)^2+u^2=0\label{eq12}.
\end{equation}
It is very difficult to solve the action $I$ for $I(t, r, \theta, \phi)$. Considering the properties of black hole spacetime, the separation of variables can be taken as follows
\begin{equation}
I=-\omega t+R(r)+H(\theta)+j\psi\label{eq13},
\end{equation}
where $\omega$ and $j$ are respectively the energy and angular momentum of the particle. Since SAdS black hole is nonrotating, the angular velocity of the particle at the horizon is $\Omega_+=\frac{d\varphi}{dt}\Bigg |_{r=r_+}=0$.
Using Eq.(\ref{eq13}) into Eq. (\ref{eq12}) and solving $R(r)$ yields an expression of
\begin{eqnarray}
R(r)=\pm\frac{r^2_+}{\Delta_{,r}(r_+)}\int \frac{dr}{(r-r_+)}\quad \times \sqrt{\omega^2-\frac{\Delta_{,r}(r_+)(r-r_+)}{r^2_+}\left[g^{22}(\partial_\theta H(\theta))^2+g^{33}j^2+u^2\right]}\label{eq14}.
\end{eqnarray}
We consider the emitted particle as an ellipsoid shell of energy  to tunnel across the event horizon and should not have  motion in  $\theta$-direction ($d\theta=0$) and therefore, finishing the above integral we get
\begin{eqnarray}
R(r)&=&\pm \frac{\pi i r^2_+}{\Delta_{,r}(r_+)}\omega+\xi \nonumber\\
&=& \frac{{\rm i} 4\pi m^2}{(r_+-m+2\frac{r^3_+}{\ell^2})}\left(1-\frac{4m^2}{\ell^2}+....\right)^2\omega +\xi\label{eq15},
\end{eqnarray}
where $\pm$ sign comes from the square root and $\xi$ is the constant of integration. Inserting Eq. (\ref{eq15}) into Eq. (\ref{eq13}), the imaginary part of actions corresponding to outgoing and incoming particles can be written as
\begin{eqnarray}
{\rm Im}I_\pm =\frac{ 4\pi m^2}{(r_+-m+2\frac{r^3_+}{\ell^2})}\left(1-\frac{4m^2}{\ell^2}+....\right)^2\omega +\xi\label{eq16}.
\end{eqnarray}
According to the classical limit given in Ref. \cite{fifty seven}, we ensure that the incoming probability to be unity when there is no refection i.e., every thing is absorbed by the horizon. In this situation the appropriate value of $\xi$ instead of zero or infinity can be taken as $\xi=\frac{ 4\pi m^2}{(r_+-m+2\frac{r^3_+}{\ell^2})}\left(1-\frac{4m^2}{\ell^2}+....\right)^2\omega+{\rm Re}(\xi)$. Therefore, ${\rm Im}I_-=0$ and $I_+$ give the imaginary part of action $I$ corresponding to the outgoing particle with the help of Eq. (\ref{eq10}) to the form
\begin{eqnarray}
{\rm Im}I =\frac{ 4\pi m^2}{(r_+-m+2\frac{r^3_+}{\ell^2})}\left(1-\frac{4m^2}{\ell^2}+....\right)^2\omega \label{eq17}.
\end{eqnarray}
Substituting Eq. (\ref{eq5}) into Eq. (\ref{eq17}), the imaginary part of action takes the form
\begin{eqnarray}
{\rm Im}I=\frac{4\pi m^2\left(1-\frac{4m^2}{\ell^2}+\cdot
\cdot\cdot\right)^2}{2m\left(1-\frac{4m^2}{\ell^2}+\cdot
\cdot\cdot\right)-m+\frac{2}{\ell^2}\left\{2m\left(1-\frac{4m^2}{\ell^2}+\cdot
\cdot\cdot\right)\right\}^3}\omega\label{eq16}\label{eq18}.
\end{eqnarray}
Since the SAdS spacetime is dynamic due to the presence of cosmological constant, we fix the Amowitt-Deser-Misner (ADM) mass of the total spacetime and allow to fluctuate. When a particle with energy $\omega$ tunnels out, the mass of the SAdS black hole changed into $m-\omega$. Since the angular velocity of the particle at the horizon is zero $(\Omega_+=0)$, the angular momentum is equal to zero. Taking the self-gravitational interaction into account, the imaginary part of the true action can be calculated from Eq. (\ref{eq18}) in the following integral form
\begin{eqnarray}
{\rm Im}I=4\pi\int^\omega_0\frac{m^2\left(1-\frac{4m^2}{\ell^2}+\cdot
\cdot\cdot\right)^2}{2m\left(1-\frac{4m^2}{\ell^2}+\cdot
\cdot\cdot\right)-m+\frac{2}{\ell^2}\{2m\left(1-\frac{4m^2}{\ell^2}+\cdot
\cdot\cdot\right)\}^3}d\omega'\label{eq19}.
\end{eqnarray}
Replacing $m$ by $m-\omega$, we have
\begin{eqnarray}
{\rm Im}I=-4\pi\int^{(m-\omega)}_m \frac{(m-\omega')^2\left(1-\frac{4(m-\omega')^2}{\ell^2}+\cdot
\cdot\right)^2}{2(m-\omega')\left(1-\frac{4(m-\omega')^2}{\ell^2}+\cdot
\cdot\right)-(m-\omega')+\frac{2}{\ell^2}\{2(m-\omega')\left(1-\frac{4(m-\omega')^2}{\ell^2}+\cdot
\cdot\right)\}^3}\nonumber\\
\times d(m-\omega')\label{eq20}.
\end{eqnarray}
Employing WKB approximation, we neglect the terms $(m-\omega')^n$ for $n\ge 5 $, and rewrite Eq. (\ref{eq20}) as
\begin{eqnarray}
{\rm Im}I&=&-4\pi\int^{(m-\omega)}_m \frac{(m-\omega')\left(1-\frac{8(m-\omega')^2}{\ell^2}\right)}{\left(1+\frac{8(m-\omega')^2}{\ell^2}\right)}\times d(m-\omega'),\nonumber\\
&=&-\frac{\pi}{2}\left[4(m-\omega)^2\left(1-\frac{8(m-\omega)^2}{\ell^2}\right)-4m^2\left(1-\frac{4m^2}{\ell^2}\right)\right]\label{eq21}.
\end{eqnarray}
Therefore, from Eq. (\ref{eq7}) the tunneling probability for SAdS black hole is given by
\begin{eqnarray}
\Gamma \sim {\rm exp}(-2{\rm Im}I)&=&{\rm exp}\left\{\pi\left[4(m-\omega)^2\left(1-\frac{8(m-\omega)^2}{\ell^2}\right)-4m^2\left(1-\frac{4m^2}{\ell^2}\right)\right]\right\}\nonumber\\
&=&{\rm exp}[\pi(r^2_f-r^2_i)]={\rm exp}(\Delta S_{BH}).\label{eq22}
\end{eqnarray}
Where, $r_f=2(m-\omega)\left(1-\frac{4(m-\omega)^2}{\ell^2}\right)$ and $r_i=2m\left(1-\frac{4m^2}{\ell^2}\right)$
are the locations of the SAdS event horizon before and after the particle emission, and $\Delta S_{BH}=S_{BH}(m-\omega)-S_{BH}(m)$ is the change of Bekenstein-Hawking entropy.

\section{Purely Thermal Radiation}\label{sec4}
The radiation spectrum described by Eq. (\ref{eq22}) is not pure thermal although gives a correction to the Hawking radiation of SAdS black hole. The purely thermal spectrum can be derived from Eq. (\ref{eq22}) by expanding the tunneling rate in power of $\omega$ upto second order as discussed by Liu et al. \cite{fifty} of the form
\begin{eqnarray}
\Gamma \sim {\rm exp}(\Delta S_{BH})&=&{\rm exp}\left\{-\omega \frac{\partial S_{BH}(m)}{\partial\omega}+\omega^2\frac{\partial^2 S_{BH}(m)}{\partial\omega^2}\right\}.\label{eq23}
\end{eqnarray}
It is clear from Eq. (\ref{eq22}) that
\begin{eqnarray}
S_{BH}(m-\omega)=4(m-\omega)^2\left(1-\frac{8(m-\omega)^2}{\ell^2}\right),\label{eq24}
\end{eqnarray}
 which gives
\begin{eqnarray}
\frac{\partial S_{BH}(m-\omega)}{\partial\omega}=8(m-\omega)\left(1-\frac{16(m-\omega)^2}{\ell^2}\right),\quad\frac{\partial^2 S_{BH}(m-\omega)}{\partial\omega^2}=-\left(1-\frac{48(m-\omega)^2}{\ell^2}\right),\label{eq25}
\end{eqnarray}
with $\omega=0$, the above equation takes the following simple form
\begin{eqnarray}
\frac{\partial S_{BH}(m)}{\partial\omega}=8\left(m-\frac{16m^3}{\ell^2}\right),\quad \frac{\partial^2 S_{BH}(m)}{\partial\omega^2}=-\left(1-\frac{48m^2}{\ell^2}\right).\label{eq26}
\end{eqnarray}
The purely thermal spectrum described by Eq. (\ref{eq23}) can be reduced with the help of Eq. (\ref{eq26}) of the form
\begin{eqnarray}
\Gamma \sim {\rm exp}(\Delta S_{BH})={\rm exp}\left\{-8\pi \omega \left[\left(m-\frac{16m^3}{\ell^2}\right)-\frac{\omega}{2}\left(1-\frac{48m^2}{\ell^2}\right)\right]\right\}.\label{eq27}
\end{eqnarray}
\twocolumn
\section{Concluding Remarks}\label{sec4}
In this paper, we have presented an extension of the classical tunneling framework \cite{fourty nine,fifty,fifty one} for the spherically symmetric black hole cases to deal with Hawking radiation of massive particles as tunneling process through the event horizon of SAdS black hole. By treating the background spacetime as dynamical, the energy  and the angular momentum as conservation, we have found the non-thermal and purely thermal tunneling probabilities of SAdS black hole when the particle's self-gravitation is taken into account.

The non-thermal tunneling probability of particle emission is proportional to the phase space factor depending on the initial and final entropy of the system, which implies that the emission spectrum actually deviates from perfect thermally but is in agreement with an underlying unitary theory. We therefore come to the conclusion that the actual radiation spectrum of SAdS black hole is not precisely thermal, which provides an interesting correction to Hawking pure thermal spectrum. In the limiting case, i.e., when $\Lambda =0$, our results for non-thermal and purely thermal radiation reduced to
\begin{eqnarray}
\Gamma \sim {\rm exp}(-2{\rm Im}I)={\rm exp}\left\{\pi\left[4(m-\omega)^2-4m^2\right]\right\},\label{eq28}
\end{eqnarray}
and
\begin{eqnarray}
\Gamma \sim {\rm exp}(\Delta S_{BH})={\rm exp}\{-8\pi\omega(m-\frac{\omega}{2})\}.\label{eq29}
\end{eqnarray}
These are the non-thermal and purely thermal tunneling rates of Schwarzschild black hole, where $r_i=2m$ and $r_f=2(m-\omega)$ are the positions of the event horizon of Schwarzschild black hole before and after the emission. Obviously, both the results are fully consistent with that obtained by Parikh and Wilczek \cite{nine}. \\
{\bf Acknowledgement}\\
One of the authors (MAR) thanks the Abdus Salam International Centre for Theoretical Physics (ICTP), Trieste, Italy, for giving opportunity to utilize its e-journals for research purpose.

\end{document}